\documentclass[twocolumn, secnumarabic,amssymb,amsmath,pre]{revtex4}
\usepackage{amsmath}
\usepackage{amssymb}
\usepackage{graphicx}
\usepackage{natbib}

\begin{document} 
\title{Electrostatic approximation of source-to-target mean first passage times on lattices}
\author{Anthony P.\ Roberts and Christophe P.\ Haynes}
\affiliation{School of Mathematics and Physics, The University of Queensland, Brisbane 4072, Australia}

\begin{abstract}
We demonstrate that the source to target mean first passage time (MFPT) is approximately given by the potential difference of an electrostatic problem which shows that the MFPT scales like the resistance  between the target and a distant shell. This analogy allows the asymmetry of the MFPT on non uniform lattices to be incorporated and provides a number of useful insights. For example, on transient lattices, the MFPT converges exactly to the product of the mass and the site dependent resistance between the target and a shell at infinity.
\end{abstract}

\maketitle

Problems which require calculation of the time it takes a random walker (RW) to reach one or more traps on a network arise
naturally in numerous settings~\cite{Redner} including chemical reaction rates~\cite{HavlinTrap,Balak}, transport in random media~\cite{Avra}, search algorithms, spreading of disease and other problems~\cite{Condamin2,Voituriez2}. The single most important characteristic of the density of trapping times is its average; the mean first passage time (MFPT). A variety of MFPT problems can be defined.  These include the time it takes to reach a trap if the RW is released at a random site on the network~\cite{Kozak,Agliara,Haynes,Zhang}, the time to reach any site a distance $r$ away~\cite{Avra}, or the time it takes a walker  to visit a target $x$ if it is released at a source $y$~\cite{Condamin2,Condamin}. The nature of the process to be modeled dictates the relevant MFPT. 

Recently, there has been interest in the scaling form for a variety of source to target MFPT problems on finite lattices~\cite{Condamin2, Voituriez}. Condamin {\em et al} derived the form   
\begin{equation}\label{approx}
\langle\tau(x|y)\rangle  \approx  N\left\{
\begin{array}{cc}
A + B r^{d_w-d_f} &  d_w-d_f > 0 \\
A + B\ln(r)   & d_w - d_f = 0 \\
A - B r^{d_w-d_f}  & d_w-d_f < 0.\\
\end{array}\right.
\end{equation}
for lattices with $N$ sites. Here $d_w$ and $d_f$ are the random walk and fractal dimensions respectively.  Interestingly $A$ and $B$ are independent of the lattice shape. The approximation provides an extremely simple solution to a problem which one might expect to show a strong dependence on boundary shape and the source to target positions. 

Although Eq.~\eqref{approx} was found to give very good predictions on a range of lattices, its derivation assumes that the diffusion propagator at all points of the lattice is identical. This is only strictly true on ``uniform'' networks such as $n$ dimensional lattices.  Recent theoretical interest in the MFPT has been stimulated by the study of transport on complex networks~\cite{Albert, Gallos} which have a high degree of non-uniformity in their coordination number and exhibit modularity. Moreover, there are numerous loopless networks which have non-uniform coordination numbers and branch lengths. The absence of loops is known to significantly affect their dynamics~\cite{Baron}. 

It is therefore interesting to consider $\langle\tau(x|y)\rangle$ on non-uniform lattices in more detail. We do this by deriving a scaling form for the MFPT using an analogous electrostatic problem.  Although our approach holds in general, we focus in particular on loopless networks and on classes of trees that do not follow the standard propagator form~\cite{HRgfe09}. The method is fruitful in the following respects; it incorporates the point to point character of the MFPT, leads to a more general scaling law, explains independence of the scaled MFPT ($\langle\tau(x|y) \rangle/N$) on lattice shape, and shows how $\tau(x,y)$ is related to resistance. 

On a finite lattice it is possible to show that the MFPT $\tau(x|y)$ for a walker released at $y$ to arrive at $x$ is exactly given by~\cite{Noh,Condamin2}
\begin{equation}\label{exact}
 \tau(x|y) = 2M\int_0^\infty \frac{p(x|x,t)}{\kappa_x} - \frac{p(y|x,t)}{\kappa_y} dt.
\end{equation}
Here $p(y|x,t)$ is the probability that a walker released at $x$ at $t=0$ is at the point $y$ at time $t$ and $M$ is the number of bonds within the structure. If the lattice is uniform ($\kappa_x = \kappa_y = \kappa$) then $2M/\kappa = N$. The probability satisfies the discrete diffusion equation $p_t= \nabla^2 p + \delta_{t,0}$ with reflective boundary conditions which has steady state solution $p(y|x,t)=\kappa_y/(2M)$. To obtain $\tau(x|y)$ exactly, $p$ has to be calculated for each finite lattice. 
In Ref.~\cite{Condamin2} the authors derive Eq.~\eqref{approx} based on the hypothesis that $p(x|x,t)/\kappa_x-p(y|x,t)/\kappa_y \approx p_\infty (y|x,t)/\kappa_x-p_\infty(y|x,t)/\kappa_y $
where $p_\infty$ is the Green's function for the case where the lattice has infinite size. Certainly both expressions are identical for small times (before the rebound from the boundary) and presumably have similar long time behavior.
Using a well known propagator form~\cite{ProcPRL,Roman} $p(x|y,t) \approx \Pi(r^{d_w}/t)/t^{d_f/d_w}$ the integrals are evaluated directly, giving rise to Eq.~\eqref{approx}.  As $\tau(x|y)$ has a point to point character, the scaling law given in Eq.~\eqref{approx} (which only depends on distance) refers to the MFPT to reach a target averaged over all sources a fixed distance away. We denote this averaging by $\langle \cdot \rangle$.

Instead of proceeding as above we show how the integrals in Eq.~\eqref{exact} can be interpreted as electrostatic potentials.
To see this consider terminating the integration at a time scale associated with diffusion of the finite lattice $T=bL^{d_w}$ where $L$ is a typical chemical length (i.e. stepping distance) on the structure and $d_w$ is the random walk dimension in chemical space. In Ref.~\cite{HRgfe09} (also see Ref.~\cite{Cates}) it is argued that the integral $P(y|x,T)\equiv \int_0^T p(y|x,t) dt \approx \kappa_y\phi(y|x;L)$,  where $\phi$ is the potential at $y$ due to the injection of a unit current at $x$ with a Dirichlet boundary condition applied at $L$. $P(y|x,T)$ exactly represent the concentration field created by the release of a walker at $x$ at each of $T+1$ time steps. The identification with the potential follows from the hypothesis that $P(y|x,T)$ has equilibrated near the origin (thus satisfying the potential equation), and that $P(y|x,T)\approx 0$ on the boundary. The latter claim is plausible because only walkers released at small times will have reached $d(x,y)=L$, where $d(x,y)$ is the chemical distance between $x$ and $y$.

Thus the MFPT is related to the potential by the expression
\begin{equation}\label{deltaphi}
\tau(x|y) \approx 2M\Delta \phi = 2M[\phi(x|x;L)-\phi(y|x;L)].
\end{equation}
To proceed recall that the point-to-shell resistance $R_x(\ell)$ is defined as the potential difference induced by the flow of unit current from the target $x$ to a grounded shell at a chemical distance $\ell$. If the potential depends on the chemical distance $\ell$ alone then $\langle \Delta \phi \rangle = R_x(\ell)$ which should be independent of $L$ (the location of the Dirichlet condition). In general the potential depends on position $y$, but not too strongly for lattices which follow the fractal Einstein law~\cite{Telcs2,HRgfe09}. Thus, we find the approximation 
\begin{equation}\label{big}
\langle \tau(x|y) \rangle  \approx 2M R_x(\ell).
\end{equation}
From the asymptotic form of the resistance for large $\ell$~\cite{Hughes2}, the MFPT (for $\ell \to \infty$) will scale as 
\begin{equation}\label{resistanceform} 
\langle \tau(x|y) \rangle \approx  2M\left\{
\begin{array}{cc}
B \ell^{\zeta} &  \zeta>0 \\
B\ln(\ell)   & \zeta=0 \\
R^*_x  - B \ell^{\zeta}  & \zeta<0 .\\
\end{array}\right.
\end{equation}
Here $R^*_x$ is the resistance from $x$ to a shell at infinity and $\zeta$ is the resistance exponent. 
The resistance $R_x(\ell)$ is not expected to strongly depend on the size of the lattice as it is dominated by the shortest path(s) between $x$ and $y$. Equation~\eqref{resistanceform} therefore provides an important insight into two striking features of the MFPT identified in Ref.~\cite{Condamin2}; the near linear dependence of the MFPT on mass, and the independence of the coefficients in Eq.~\eqref{approx} on the shape of the confining domain. 

If the fractal Einstein law $\zeta = d_w-d_f$ is assumed, then the argument recovers the scaling form of Eq.~\eqref{approx} except that the coefficients are now determined by the electrostatic problem. The electrostatic scaling forms (which are strictly true only for large $\ell$) give important insights into the constants $A$ and $B$ appearing in Eq.~\eqref{approx}. For $\zeta \geq 0$, there is strictly no constant in the asymptotic form while for $\zeta < 0$, $NA = 2MR^*_x$. Note that there are lattices on which  $\zeta \neq d_w - d_f$, such as those created by diffusion limited aggregation~\cite{Witten} (DLA) and certain asymmetric deterministic trees~\cite{HRgfe09}. Thus the resistance form expressed in terms of $\zeta$ is more general.  

The MFPT is also connected to the point-to-point resistivity $\rho(x,y)$ which can be defined as the potential difference between two sites if a unit current is injected at one and withdrawn at the other. If the Laplacian of Eq.~\eqref{exact} is taken and the right hand side is integrated,  $\tau(x|y)$ can be shown to satisfy $\nabla^2_y \tau(x|y) = 1/M-\delta_{xy}$~\cite{Condamin2}. Now consider the equation satisfied by the function
$\chi(z)= [\tau(y|z) - \tau(x|z)]/(2M)$; $\nabla^2_z \chi = \delta_{xz}-\delta_{yz}$ with reflective boundary
conditions. The latter equation is exactly the potential associated with the flow of unit current from $x$ to $y$ so $\rho(x,y)=\chi(x)-\chi(y)$ which recovers a result of Chandra~\cite{Chandra}
\begin{equation}\label{Chandra}
\tau(x|y) + \tau(y|x) = 2M\rho(x,y).
\end{equation}
Note that if $\tau(x|y) = \tau(y|x)$ then the MFPT is exactly given by $\tau(x|y) = M\rho(x,y)$~\cite{Gefen} (see \cite{Hilfer} for a point to shell analogue). This is obviously true on uniform networks in which every site is identical (e.g. $n$-$d$ lattices). 

In terms of current discussion, the fact that $\tau(x|y) = M\rho(x,y)$ on uniform lattices offers two interesting insights.
First, we have argued that $\tau(x|y) \approx 2M\Delta\phi$ which implies that $\rho(x,y)\approx 2\Delta\phi$. This result becomes exact for infinite uniform lattices and is in fact useful for calculating $\rho(x,y)$~\cite{Cserti}.
Secondly, it allows a useful interpretation of the MFPT derivation given in Ref.~\cite{Condamin2}. For uniform finite lattices Eq.\eqref{exact} becomes
$
\rho(x,y) = \int_0^\infty p(x|x,t) - p(y|x,t) dt
$
which should remain well defined for infinite lattices giving 
$
\rho_\infty(x,y) = \int_0^\infty p_\infty(x|x,t) - p_\infty(y|x,t) dt.
$
Thus the finite to infinite assumption can be viewed as $\tau(x|y) = M\rho(x,y) \approx M\rho_\infty(x,y)$ which coincides with the form \eqref{big}. 

It is generally not clear how good an approximation $\tau(x|y) = \tau(y|x)$ is for non-uniform lattices. 
Certainly it will hold exactly for pairs of points which are reflections in lines of symmetries of the structure (e.g. the Sierpinski gasket).  An obvious counter example is if $x$ is the end of a dead end branch and $y$ is its neighbor. In this case $\tau(x|y) = 1$ whereas $\tau(y|x) = 2M - 1$ since $\rho(x,y) = 1$. An expression for the asymmetric component of the MFPT
is found by combining Eqs.~\eqref{exact} and~\eqref{Chandra} and using the identity
 $\kappa_y p(x|y,t)= \kappa_x p(y|x,t)$ to obtain
\begin{equation}\label{eq:final8}
\tau(x|y)  = M\rho(x,y) + M\int_0^\infty \frac{p(x|x,t)}{\kappa_x}- \frac{p(y|y,t)}{\kappa_y} dt.
\end{equation} 
Note that the integral term of Eq.~\eqref{eq:final8} has been studied in detail by Noh and Rieger~\cite{Noh}. 
The formula is useful below, and also clearly demonstrates how the assumption of a site independent
form of the propagator $p(y|x,t)/\kappa_y$ is connected to symmetric MFPTs.

Benichou et al.~\cite{Voituriez} give an exact expression for the average of $\tau(x|y)$ over the $\kappa_x$ neighbors of $x$ as $\langle \tau(1)\rangle  = 2M/\kappa_x -1$. From Eq.~\eqref{big}, this implies that the point to shell resistance $R_x(1) \approx \langle \tau(1)\rangle /(2M) \approx 1/\kappa_x$. The approximation is actually exact because there are $\kappa_x$ resistors in parallel between the site $x$ and the shell grounded at $d(x,y) = 1$. In Ref.~\cite{Voituriez} it was hypothesized that the constants $A$ and $B$ could be evaluated by using the boundary condition $\langle \tau(1)\rangle/N \approx 1$ (these were termed 'zero constant' formulas~\cite{Voituriez}). If a similar analysis is performed for non-uniform lattices this would imply $B = 1/\kappa_x$ for $\zeta >0$ and $B = R^*_x - 1/\kappa_x$ for $\zeta <0$ in Eq.~\eqref{resistanceform}. Note that the form of the resistance for large $\ell$ is asymptotic in nature and there is no guarantee that extrapolating the scaling form to $\ell=1$ to fix the constants will be accurate. We test the results below. 

If the MFPT is averaged over all sources and targets a chemical distance $\ell$ apart then $\bar{\tau}(\ell) = M\bar{\rho}(\ell)$ exactly. Here $\bar{\rho}(\ell)$ is defined by the same averaging process (recall $\rho$ is the point to point resistance). This can easily be seen from Eq.~\eqref{Chandra} or \eqref{eq:final8}. For loopless networks $\rho(\ell) = \ell$ and therefore $\bar{\tau}(\ell) = M\ell$ exactly. This form is confirmed by the numerical results in Fig.\ 2(b) of Ref.~\cite{Voituriez} for the $T$-tree. However it is not clear that the coefficients of $\bar{\tau}({\ell})$  and $\langle \tau(x|y)\rangle$ are identical. Furthermore, the scaling exponents of averaged quantities may differ from that of the original quantity~\cite{Burioni5}. Indeed, for loopless network having $\xi <0$, the form $\langle \tau(x|y)\rangle \sim 2M(R^*_x - B\ell^{\zeta})$ does not scale as $M\ell$.

\begin{figure}
\begin{center}
\includegraphics[width = 0.3\textwidth]{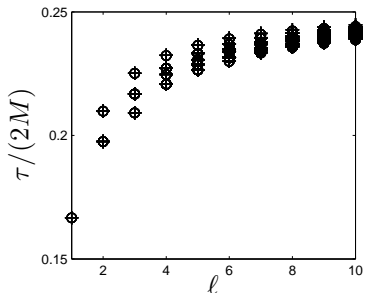}
\end{center}
\vspace{-5mm}

\caption{The MFPT on a finite 3-$d$ network ($\circ$) is almost exactly given by $2\Delta \phi$ ($+$) on an ``infinite"  lattice}
\vspace{-5mm}
\label{cube}
\end{figure}

To investigate the approximation $\tau \approx 2M\Delta \phi$ we have calculated the MFPT from each source on a range of finite lattices to a fixed target $x$. The MFPT $\tau(x|y)$ is computed by releasing a walker at the target $x$ and using the resulting $p(y|x,t)$ to evaluate the integrals at each $y$ of Eq.~\eqref{exact} exactly. The question is whether the potential difference $\Delta \phi$ on an ``infinite'' lattice can give a good approximation to the MFPT on finite sub-domains around the target. 
We first consider the MFPT on a finite 3-$d$ lattice $\ell\leq 15$ with a target at its center. The potential $\Delta \phi$ is computed on a 3-$d$ lattice with Dirichlet boundary conditions at $L = 50$. In this case $\Delta \phi$ (Fig.~\ref{cube}) reproduces the point to point MFPT almost exactly. 
\begin{figure}
\begin{center}
\includegraphics[width = 0.48\textwidth]{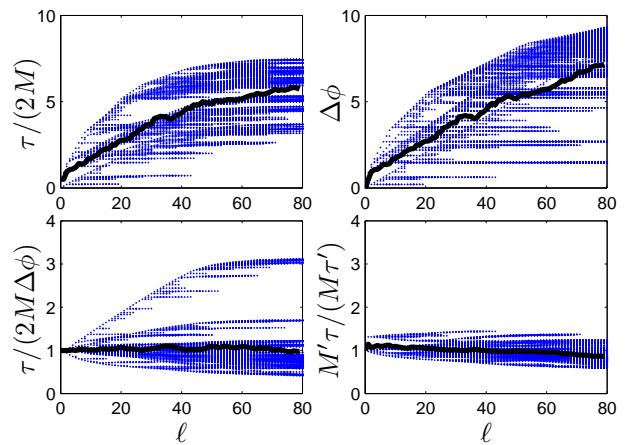}
\end{center}
\vspace{-7mm}

\caption{The MFPT of a DLA cluster clipped at $\ell=80$ (top left) compared with the potential on the same DLA cluster with Dirichlet conditions applied at $\ell=140$ (top right). The ratio of the two functions is shown bottom left. The final figure (bottom right) compares the scaled MFPTs  obtained for the cluster clipped at radii $\ell = 80$ \& $\ell = 140$.}
\vspace{-5mm}

\label{DLA}
\end{figure}

\begin{figure}
\begin{center}
\includegraphics[width = 0.25\textwidth]{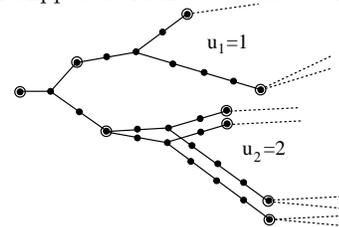}
\end{center}
\vspace{-7mm}

\caption{An example of a deterministic tree (DT) constructed by attaching copies of the rescaled base unit to each end of the base unit's end points and so forth~\cite{HRgfe09,arxivasym}. Shown is DT-C which has $(u_1,u_2) = (1,2)$. We also consider DT-A which has $(u_1,u_2) = (2,1)$ and DT-B which has $(u_1,u_2) = (1,1)$ but the length of successive iterates quadruples.}
\vspace{-5mm}
\label{treetest3}
\end{figure}

\begin{figure}
\begin{center}
\includegraphics[width = 0.48\textwidth]{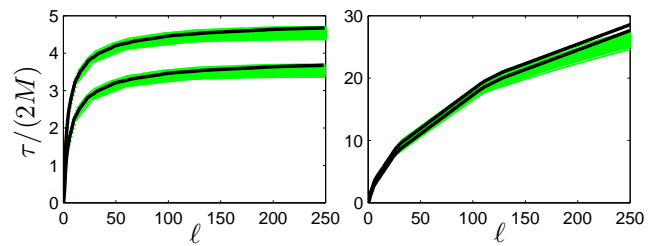}
\end{center}
\vspace{-5mm}

\caption{The MFPT for DT-A (left) ($\zeta = -0.58$) for two different targets. The solid lines are $\langle \nabla \phi \rangle$ and confirm that $\tau(x|y) \approx 2MR_x(\ell)$ with the limiting values given by $R^*_x$. The results for the model DT-B ($\zeta = 1/2$) are shown on the right. For these models $\zeta = d_w - d_f$.}
\vspace{-5mm}

\label{treetest}
\end{figure}

\begin{figure}
\begin{center}
\includegraphics[width = 0.48\textwidth]{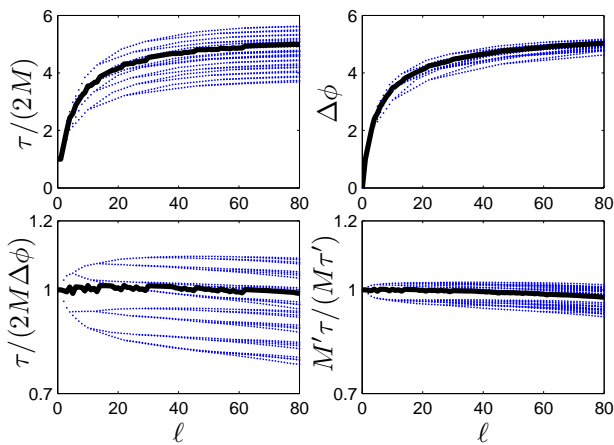}
\end{center}
\vspace{-7mm}

\caption{Results for model DT-C. The sub-plots match those of Fig~\ref{DLA}. For this lattice $\zeta \neq d_w - d_f$ implying highly anisotropic transport.}
\vspace{-5mm}

\label{treetest2}
\end{figure}
\begin{figure}
\begin{center}
\includegraphics[width = 0.48\textwidth]{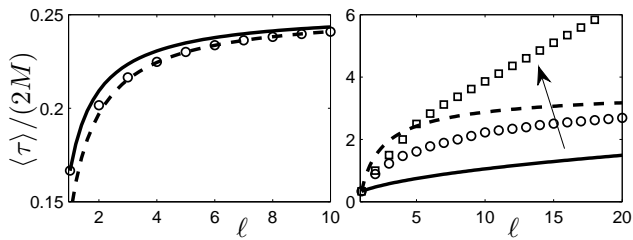}
\end{center}
\vspace{-5mm}

\caption{A comparison of $\langle \tau \rangle$ ($\circ$) to the zero-constant formula described in the text.
In the left figure the dashed line is the asymptotically fitted form. The right figure
shows the results for networks DT-B ($\square$, solid line) and DT-A ($\circ$, dashed line).}
   
\vspace{-5mm}
\label{cubetest}
\end{figure}

The MFPT of a DLA cluster is shown in Fig.~\ref{DLA}.
 It is seen that $\Delta \phi$ captures many features of $\tau(x|y)$. In particular, the MFPT is seen to be nearly constant along dead end branches with value given by the MFPT at each branching point. This exactly matches the qualitative behavior of the potential. The solid line in the lower left box shows that $\langle\tau \rangle/(2M\langle\Delta \phi \rangle) \approx 1$ (supporting the scaling form~\eqref{resistanceform}) but the point to point comparison begins to show significant outliers. The graph on the lower right shows that a practical approximation is obtained from $\tau \approx M/M' \times \tau^\prime$ where the prime indicates a larger lattice. This is expected, and indicates that the MFPT on a lattices of size $M<M^\prime$ can be estimated accurately by multiplying $\tau^\prime$ by a simple factor. 

We now consider the three deterministic trees (DT-A,B,C) described in Fig.~\ref{treetest3}. The MFPT for DT-A ($\zeta=-0.58$) is shown in Fig.~\ref{treetest} for the cases of two targets; one located at base of the tree and the second at its immediate neighbor.  The MFPT exhibits mild anisotropy, but is very well quantitatively approximated by $\langle\Delta\phi\rangle$ on a larger lattice.
A similar picture is shown for tree DT-B which has $\zeta=\frac12$. The properties of both networks DT-A and DT-C obey
the FE law $\zeta=d_w-d_f$ so follow the scaling of Eq.~\eqref{approx} (and therefore Eq.~\eqref{resistanceform}). 
In contrast, on network DT-C, $\zeta \neq d_w-d_f$ and transport is therefore highly anisotropic. This is shown in Fig.~\ref{treetest2}.  The point to point approximation $\Delta \phi$ is only able to reproduce the MFPT at small $\ell$ for this model, but it is clear that
$\langle\tau \rangle \approx 2M\langle\Delta \phi \rangle$ (solid line, lower left box) confirming  $\langle \tau \rangle \approx 2M R_x$ and hence Eq.~\eqref{resistanceform}.

In Fig.~\ref{cubetest} we plot $\langle \tau \rangle$ at a single target vs the 'zero constant' formula with the values of $B$ described previously. On the uniform 3-$d$ lattice the formula appears reasonable but if $B$ is obtained from the large $\ell$ behavior the value is 0.11 (dashed line in Fig.~\ref{cubetest}) rather than 0.0854. For the non-uniform DT-A and DT-B networks the approximations are seen to be poor and we conclude that the 'zero-constant' approach~\cite{Voituriez} does not generally work for non-uniform lattices. 

We have shown how the source to target MFPT on a finite lattice can be interpreted as a potential difference $2M\Delta \phi$ on an infinite extension of the lattice.  The approximation $2M\Delta \phi$ reproduces the point to point character of $\tau(x|y)$ of the function near the target, particularly for uniform lattices. There is some advantage in obtaining $\tau(x|y)$ from a single calculation of $\Delta \phi$ in that an approximation for lattices of all sizes is obtained, however a significantly better approximation is yielded by $\tau(x|y) = (M/M')\tau'(x|y)$ where $\tau'(x|y)$ is the MFPT on a large lattice (bottom right panels of Figs.~\ref{DLA} and~\ref{treetest2}). 

The spatially averaged MFPT (over sources at distance $\ell$) $\langle \tau \rangle \approx 2M  \langle \Delta \phi \rangle$ is identified as the point to shell resistance $\langle \tau \rangle \approx 2MR_x(\ell)$ because the potential $\phi$ is associated with a unit current flowing between the target and a distant shell. These expressions are confirmed by the numerical results. Note that the scaling forms~\eqref{resistanceform} are more general than~\eqref{approx} in the sense that they encompass lattices which do not follow the fractal Einstein law (e.g. DLA and DT-C). Moreover, the form does not involve the fractal dimension directly, and hence we propose that it can hold for lattices with a well defined exponent $\zeta$ (e.g.\ bundled structures~\cite{Cassi1} such as combs.)

Given the connection between the MFPT and resistance, we believe that further insights may be gained into other problems (e.g. a random walker leaving a confined domain~\cite{Voituriez2}) by studying their electrostatic analogs.


\begin{thebibliography}{30}
\expandafter\ifx\csname natexlab\endcsname\relax\def\natexlab#1{#1}\fi
\expandafter\ifx\csname bibnamefont\endcsname\relax
  \def\bibnamefont#1{#1}\fi
\expandafter\ifx\csname bibfnamefont\endcsname\relax
  \def\bibfnamefont#1{#1}\fi
\expandafter\ifx\csname citenamefont\endcsname\relax
  \def\citenamefont#1{#1}\fi
\expandafter\ifx\csname url\endcsname\relax
  \def\url#1{\texttt{#1}}\fi
\expandafter\ifx\csname urlprefix\endcsname\relax\def\urlprefix{URL }\fi
\providecommand{\bibinfo}[2]{#2}
\providecommand{\eprint}[2][]{\url{#2}}

\bibitem[{\citenamefont{Redner}(2001)}]{Redner}
\bibinfo{author}{\bibfnamefont{S.}~\bibnamefont{Redner}},
  \emph{\bibinfo{title}{A guide to first passage processes}}
  (\bibinfo{publisher}{Cambridge University Press},
  \bibinfo{address}{Cambridge}, \bibinfo{year}{2001}).

\bibitem[{\citenamefont{Balakrishnan}(1995)}]{Balak}
\bibinfo{author}{\bibfnamefont{V.}~\bibnamefont{Balakrishnan}},
  \bibinfo{journal}{Mat. Sci. Eng., B} \textbf{\bibinfo{volume}{32}},
  \bibinfo{pages}{201} (\bibinfo{year}{1995}).

\bibitem[{\citenamefont{Havlin et~al.}(1991)\citenamefont{Havlin, Kopelman,
  Schoonover, and Weiss}}]{HavlinTrap}
\bibinfo{author}{\bibfnamefont{S.}~\bibnamefont{Havlin}},
  \bibinfo{author}{\bibfnamefont{R.}~\bibnamefont{Kopelman}},
  \bibinfo{author}{\bibfnamefont{R.}~\bibnamefont{Schoonover}},
  \bibnamefont{and} \bibinfo{author}{\bibfnamefont{G.~H.} \bibnamefont{Weiss}},
  \bibinfo{journal}{Phys. Rev. A} \textbf{\bibinfo{volume}{43}},
  \bibinfo{pages}{5228} (\bibinfo{year}{1991}).

\bibitem[{\citenamefont{ben Avraham and Havlin}(2000)}]{Avra}
\bibinfo{author}{\bibfnamefont{D.}~\bibnamefont{ben Avraham}} \bibnamefont{and}
  \bibinfo{author}{\bibfnamefont{S.}~\bibnamefont{Havlin}},
  \emph{\bibinfo{title}{Diffusion and Reactions in Fractals and Disordered
  Systems}} (\bibinfo{publisher}{Cambridge Univ. Press},
  \bibinfo{address}{Cambridge, UK}, \bibinfo{year}{2000}).

\bibitem[{\citenamefont{Condamin
  et~al.}(2007{\natexlab{a}})\citenamefont{Condamin, Benichou, Tejedor,
  Voituriez, and Klafter}}]{Condamin2}
\bibinfo{author}{\bibfnamefont{S.}~\bibnamefont{Condamin}},
  \bibinfo{author}{\bibfnamefont{O.}~\bibnamefont{Benichou}},
  \bibinfo{author}{\bibfnamefont{V.}~\bibnamefont{Tejedor}},
  \bibinfo{author}{\bibfnamefont{R.}~\bibnamefont{Voituriez}},
  \bibnamefont{and} \bibinfo{author}{\bibfnamefont{J.}~\bibnamefont{Klafter}},
  \bibinfo{journal}{Nature} \textbf{\bibinfo{volume}{450}}, \bibinfo{pages}{77}
  (\bibinfo{year}{2007}{\natexlab{a}}).

\bibitem[{\citenamefont{Benichou and Voituriez}(2008)}]{Voituriez2}
\bibinfo{author}{\bibfnamefont{O.}~\bibnamefont{Benichou}} \bibnamefont{and}
  \bibinfo{author}{\bibfnamefont{R.}~\bibnamefont{Voituriez}},
  \bibinfo{journal}{Phys. Rev. Lett.} \textbf{\bibinfo{volume}{100}},
  \bibinfo{pages}{168105} (\bibinfo{year}{2008}).

\bibitem[{\citenamefont{Agliari}(2008)}]{Agliara}
\bibinfo{author}{\bibfnamefont{E.}~\bibnamefont{Agliari}},
  \bibinfo{journal}{Phys. Rev. E} \textbf{\bibinfo{volume}{77}},
  \bibinfo{pages}{011128} (\bibinfo{year}{2008}).

\bibitem[{\citenamefont{Haynes and Roberts}(2008)}]{Haynes}
\bibinfo{author}{\bibfnamefont{C.~P.} \bibnamefont{Haynes}} \bibnamefont{and}
  \bibinfo{author}{\bibfnamefont{A.~P.} \bibnamefont{Roberts}},
  \bibinfo{journal}{Phys. Rev. E} \textbf{\bibinfo{volume}{78}},
  \bibinfo{pages}{041111} (\bibinfo{year}{2008}).

\bibitem[{\citenamefont{Zhang et~al.}(2009)\citenamefont{Zhang, Qi, Zhou, Xie,
  and Guan}}]{Zhang}
\bibinfo{author}{\bibfnamefont{Z.}~\bibnamefont{Zhang}},
  \bibinfo{author}{\bibfnamefont{Y.}~\bibnamefont{Qi}},
  \bibinfo{author}{\bibfnamefont{S.}~\bibnamefont{Zhou}},
  \bibinfo{author}{\bibfnamefont{W.}~\bibnamefont{Xie}}, \bibnamefont{and}
  \bibinfo{author}{\bibfnamefont{J.}~\bibnamefont{Guan}},
  \bibinfo{journal}{Phys. Rev. E} \textbf{\bibinfo{volume}{79}},
  \bibinfo{pages}{021127} (\bibinfo{year}{2009}).

\bibitem[{\citenamefont{Kozak and Balakrishnan}(2002)}]{Kozak}
\bibinfo{author}{\bibfnamefont{J.~J.} \bibnamefont{Kozak}} \bibnamefont{and}
  \bibinfo{author}{\bibfnamefont{V.}~\bibnamefont{Balakrishnan}},
  \bibinfo{journal}{Int. J. Bifurc. Chaos} \textbf{\bibinfo{volume}{12}},
  \bibinfo{pages}{2379} (\bibinfo{year}{2002}).

\bibitem[{\citenamefont{Condamin
  et~al.}(2007{\natexlab{b}})\citenamefont{Condamin, Benichou, and
  Moreau}}]{Condamin}
\bibinfo{author}{\bibfnamefont{S.}~\bibnamefont{Condamin}},
  \bibinfo{author}{\bibfnamefont{O.}~\bibnamefont{Benichou}}, \bibnamefont{and}
  \bibinfo{author}{\bibfnamefont{M.}~\bibnamefont{Moreau}},
  \bibinfo{journal}{Phys. Rev. E} \textbf{\bibinfo{volume}{75}},
  \bibinfo{pages}{021111} (\bibinfo{year}{2007}{\natexlab{b}}).

\bibitem[{\citenamefont{Benichou et~al.}(2008)\citenamefont{Benichou, Meyer,
  Tejedor, and Voituriez}}]{Voituriez}
\bibinfo{author}{\bibfnamefont{O.}~\bibnamefont{Benichou}},
  \bibinfo{author}{\bibfnamefont{B.}~\bibnamefont{Meyer}},
  \bibinfo{author}{\bibfnamefont{V.}~\bibnamefont{Tejedor}}, \bibnamefont{and}
  \bibinfo{author}{\bibfnamefont{R.}~\bibnamefont{Voituriez}},
  \bibinfo{journal}{Phys. Rev. Lett.} \textbf{\bibinfo{volume}{101}},
  \bibinfo{pages}{130601} (\bibinfo{year}{2008}).

\bibitem[{\citenamefont{Albert and Barab$\acute{a}$si}(2002)}]{Albert}
\bibinfo{author}{\bibfnamefont{R.}~\bibnamefont{Albert}} \bibnamefont{and}
  \bibinfo{author}{\bibfnamefont{A.~L.} \bibnamefont{Barab$\acute{a}$si}},
  \bibinfo{journal}{Rev. Mod. Phys.} \textbf{\bibinfo{volume}{74}},
  \bibinfo{pages}{47} (\bibinfo{year}{2002}).

\bibitem[{\citenamefont{Gallos et~al.}(2007)\citenamefont{Gallos, Song, Havlin,
  and Makse}}]{Gallos}
\bibinfo{author}{\bibfnamefont{L.~K.} \bibnamefont{Gallos}},
  \bibinfo{author}{\bibfnamefont{C.}~\bibnamefont{Song}},
  \bibinfo{author}{\bibfnamefont{S.}~\bibnamefont{Havlin}}, \bibnamefont{and}
  \bibinfo{author}{\bibfnamefont{H.~A.} \bibnamefont{Makse}},
  \bibinfo{journal}{Proc. Natl. Acad. Sci} \textbf{\bibinfo{volume}{104}},
  \bibinfo{pages}{7746} (\bibinfo{year}{2007}).

\bibitem[{\citenamefont{Baronchelli et~al.}(2008)\citenamefont{Baronchelli,
  Catanzaro, and Pastor-Satorras}}]{Baron}
\bibinfo{author}{\bibfnamefont{A.}~\bibnamefont{Baronchelli}},
  \bibinfo{author}{\bibfnamefont{M.}~\bibnamefont{Catanzaro}},
  \bibnamefont{and}
  \bibinfo{author}{\bibfnamefont{R.}~\bibnamefont{Pastor-Satorras}},
  \bibinfo{journal}{Phys. Rev. E} \textbf{\bibinfo{volume}{78}},
  \bibinfo{pages}{011114} (\bibinfo{year}{2008}).

\bibitem[{\citenamefont{Haynes and Roberts}(2009)}]{HRgfe09}
\bibinfo{author}{\bibfnamefont{C.~P.} \bibnamefont{Haynes}} \bibnamefont{and}
  \bibinfo{author}{\bibfnamefont{A.~P.} \bibnamefont{Roberts}},
  \bibinfo{journal}{Phys. Rev. Lett.} \textbf{\bibinfo{volume}{103}},
  \bibinfo{pages}{020601} (\bibinfo{year}{2009}).

\bibitem[{\citenamefont{Noh and Rieger}(2004)}]{Noh}
\bibinfo{author}{\bibfnamefont{J.~D.} \bibnamefont{Noh}} \bibnamefont{and}
  \bibinfo{author}{\bibfnamefont{H.}~\bibnamefont{Rieger}},
  \bibinfo{journal}{Phys. Rev. Lett.} \textbf{\bibinfo{volume}{92}},
  \bibinfo{pages}{118701} (\bibinfo{year}{2004}).

\bibitem[{\citenamefont{O'Shaughnessy and Procaccia}(1985)}]{ProcPRL}
\bibinfo{author}{\bibfnamefont{B.}~\bibnamefont{O'Shaughnessy}}
  \bibnamefont{and}
  \bibinfo{author}{\bibfnamefont{I.}~\bibnamefont{Procaccia}},
  \bibinfo{journal}{Phys. Rev. Lett.} \textbf{\bibinfo{volume}{54}},
  \bibinfo{pages}{455} (\bibinfo{year}{1985}).

\bibitem[{\citenamefont{Roman}(2004)}]{Roman}
\bibinfo{author}{\bibfnamefont{H.~E.} \bibnamefont{Roman}},
  \bibinfo{journal}{Fractals} \textbf{\bibinfo{volume}{12}},
  \bibinfo{pages}{149} (\bibinfo{year}{2004}).

\bibitem[{\citenamefont{Cates}(1985)}]{Cates}
\bibinfo{author}{\bibfnamefont{M.~E.} \bibnamefont{Cates}},
  \bibinfo{journal}{Phys. Rev. Lett.} \textbf{\bibinfo{volume}{55}},
  \bibinfo{pages}{131} (\bibinfo{year}{1985}).

\bibitem[{\citenamefont{Telcs}(1989)}]{Telcs2}
\bibinfo{author}{\bibfnamefont{A.}~\bibnamefont{Telcs}},
  \bibinfo{journal}{Probab. Th. Rel. Fields} \textbf{\bibinfo{volume}{82}},
  \bibinfo{pages}{435} (\bibinfo{year}{1989}).

\bibitem[{\citenamefont{Hughes}(1996)}]{Hughes2}
\bibinfo{author}{\bibfnamefont{B.~D.} \bibnamefont{Hughes}},
  \emph{\bibinfo{title}{Random walks and Random Environments Volume 2}}
  (\bibinfo{publisher}{Clarendon Press}, \bibinfo{address}{Oxford},
  \bibinfo{year}{1996}).

\bibitem[{\citenamefont{Witten and Sander}(1981)}]{Witten}
\bibinfo{author}{\bibfnamefont{T.~A.} \bibnamefont{Witten}} \bibnamefont{and}
  \bibinfo{author}{\bibfnamefont{L.~M.} \bibnamefont{Sander}},
  \bibinfo{journal}{Phys. Rev. Lett.} \textbf{\bibinfo{volume}{47}},
  \bibinfo{pages}{1400} (\bibinfo{year}{1981}).

\bibitem[{\citenamefont{Chandra et~al.}(1989)\citenamefont{Chandra, Raghavan,
  Ruzzo, Smolensky, and Tiwari}}]{Chandra}
\bibinfo{author}{\bibfnamefont{A.~K.} \bibnamefont{Chandra}},
  \bibinfo{author}{\bibfnamefont{P.}~\bibnamefont{Raghavan}},
  \bibinfo{author}{\bibfnamefont{W.~L.} \bibnamefont{Ruzzo}},
  \bibinfo{author}{\bibfnamefont{R.}~\bibnamefont{Smolensky}},
  \bibnamefont{and} \bibinfo{author}{\bibfnamefont{P.}~\bibnamefont{Tiwari}},
  \bibinfo{journal}{Proc. 21st ACM Symp. Th. Computing}
  \textbf{\bibinfo{volume}{.}}, \bibinfo{pages}{574} (\bibinfo{year}{1989}).

\bibitem[{\citenamefont{Gefen and Goldhirsch}(1987)}]{Gefen}
\bibinfo{author}{\bibfnamefont{Y.}~\bibnamefont{Gefen}} \bibnamefont{and}
  \bibinfo{author}{\bibfnamefont{I.}~\bibnamefont{Goldhirsch}},
  \bibinfo{journal}{Phys. Rev. B.} \textbf{\bibinfo{volume}{35}},
  \bibinfo{pages}{8639} (\bibinfo{year}{1987}).

\bibitem[{\citenamefont{Hilfer and Blumen}(1988)}]{Hilfer}
\bibinfo{author}{\bibfnamefont{R.}~\bibnamefont{Hilfer}} \bibnamefont{and}
  \bibinfo{author}{\bibfnamefont{A.}~\bibnamefont{Blumen}},
  \bibinfo{journal}{Phys. Rev. A.} \textbf{\bibinfo{volume}{37}},
  \bibinfo{pages}{578} (\bibinfo{year}{1988}).

\bibitem[{\citenamefont{Cserti}(2000)}]{Cserti}
\bibinfo{author}{\bibfnamefont{J.}~\bibnamefont{Cserti}}, \bibinfo{journal}{Am.
  J. Phys.} \textbf{\bibinfo{volume}{68}}, \bibinfo{pages}{896}
  (\bibinfo{year}{2000}).

\bibitem[{\citenamefont{Burioni and Cassi}(2005)}]{Burioni5}
\bibinfo{author}{\bibfnamefont{R.}~\bibnamefont{Burioni}} \bibnamefont{and}
  \bibinfo{author}{\bibfnamefont{D.}~\bibnamefont{Cassi}}, \bibinfo{journal}{J.
  Phys. A: Math. Gen.} \textbf{\bibinfo{volume}{38}}, \bibinfo{pages}{R45}
  (\bibinfo{year}{2005}).

\bibitem[{\citenamefont{Haynes and Roberts}(.)}]{arxivasym}
\bibinfo{author}{\bibfnamefont{C.~P.} \bibnamefont{Haynes}} \bibnamefont{and}
  \bibinfo{author}{\bibfnamefont{A.~P.} \bibnamefont{Roberts}},
  \bibinfo{journal}{arXiv:0904.3791v1} \textbf{\bibinfo{volume}{.}},
  \bibinfo{pages}{.} (\bibinfo{year}{.}).

\bibitem[{\citenamefont{Cassi and Regina}(1996)}]{Cassi1}
\bibinfo{author}{\bibfnamefont{D.}~\bibnamefont{Cassi}} \bibnamefont{and}
  \bibinfo{author}{\bibfnamefont{S.}~\bibnamefont{Regina}},
  \bibinfo{journal}{Phys. Rev. Lett.} \textbf{\bibinfo{volume}{76}},
  \bibinfo{pages}{2914} (\bibinfo{year}{1996}).

\end{thebibliography}
\end{document}